\documentclass[sn-mathphys]{sn-jnl}% Math and Physical Sciences Reference Style
\jyear{2021}%

\theoremstyle{thmstyleone}%
\theoremstyle{thmstyletwo}%

\theoremstyle{thmstylethree}%

\begin{document}

%\linenumbers

\title[Article Title]{Primary visual cortex contributes to color constancy by predicting rather than discounting the illuminant: evidence from a computational study}


\author[1]{\fnm{Shaobing} \sur{Gao}}\email{gaoshaobing@scu.edu.cn}

%\author[1]{\fnm{Zongyi} \sur{Zhan}}\email{zhan\_zongyi@163.com }

\author*[2]{\fnm{Yongjie} \sur{Li}}\email{liyj@uestc.edu.cn}

\affil[1]{\orgdiv{College of Computer Science}, \orgname{Sichuan University}, \orgaddress{
		%\street{No.24 South Section 1, Yihuan Road}, 
\city{Chengdu}, \postcode{610065}, \state{Sichuan}, \country{China}}}

%\affil[2]{\orgdiv{Department}, \orgname{Organization}, \orgaddress{\street{Street}, \city{City}, \postcode{10587}, \state{State}, \country{Country}}}

\affil*[2]{\orgdiv{School of Life Science and Technology}, \orgname{University of Electronic Science and Technology of China}, \orgaddress{
		%\street{Street}, 
\city{Chengdu}, \postcode{610054}, \state{Sichuan}, \country{China}}}


\abstract{
Color constancy (CC) is an important ability of the human visual system to stably perceive the colors of objects despite considerable changes in the color of the light illuminating them. While increasing evidence from the field of neuroscience supports that multiple levels of the visual system contribute to the realization of CC, how the primary visual cortex (V1) plays role in CC is not fully resolved. In specific, double-opponent (DO) neurons in V1 have been thought to contribute to realizing a degree of CC, but the computational mechanism is not clear. We build an electrophysiologically based V1 neural model to learn the color of the light source from a natural image dataset with the ground truth illuminants as the labels. Based on the qualitative and quantitative analysis of the responsive properties of the learned model neurons%convolution neurons of trained V1 models using singular value decomposition and classification
, we found that both the spatial structures and color weights of the receptive fields of the learned model neurons are quite similar to those of the simple and DO neurons recorded in V1. Computationally, DO cells perform more robustly than the simple cells in V1 for illuminant prediction. Therefore, this work provides computational evidence supporting that V1 DO neurons serve to realize color constancy by encoding the illuminant, which is contradictory to the common hypothesis that V1 contributes to CC by discounting the illuminant using its DO cells. This evidence is expected to not only help resolve the visual mechanisms of CC, but also provide inspiration to develop more effective computer vision models. % rather than discounting the illuminant.	

}




\maketitle

\section{Introduction}\label{sec1}

Color constancy (CC) refers to the characteristic that the human visual system (HVS) can adaptively exclude the influence of lighting conditions on the color of objects and intrinsically reflect the properties of object. {\color{black}Understanding the neural mechanism of color constancy (CC) is very important as it can provide a more sophisticated neuroscientific basis for the diagnosis and treatment of diseases related to color vision in the HVS. Furthermore, it may help us understand how the HVS processes the incoming physical light stimuli and transforms them into subjective color perception \cite{kim2020neural}. This, in turn, may inspire the field of computer vision to develop novel computational algorithms that can process dynamically varying light information in the real world. %Understanding neural mechanism of CC is very important since it can provide a more sophisticated neuroscientific basis for the diagnosis and treatment of diseases related to color vision in HSV. Furthermore, interpreting how neurons in HVS contribute to CC may help us understand how HVS processes the incoming physical light stimuli and transforms it to the subjective color perception \cite{kim2020neural}. This, in turn, may inspire novel computational algorithms that can process the dynamic varying light information in a real-world. 
}

Although many previous psychological experiments have proved the various extents to which CC affects the HVS under different control conditions and factors \cite{foster2011color,bannert2017invariance}, the neural correlations of such performance are not yet known.
%Color visual signals are initially transmitted and processed in the form of trichromatic in the retina, until the ganglion cell layer of the retina. Then, the output of retina encoded in a color-opponent manner is further projected to the lateral geniculate body (LGN), V1 and other advanced visual cortex (V2, V3, V4 areas and beyond) to proceed the next step. During the color signal processing of primate visual system, two typical types of color-sensitive neurons are clearly defined and studied \cite{shapley2011color,gegenfurtner2003cortical,solomon2007machinery}.
Electrophysiological studies have shown that the most color-sensitive neurons in the retina and LGN are the single-opponent neurons that process the red-green, blue-yellow, and black-white information \cite{shapley2011color,gegenfurtner2003cortical,solomon2007machinery}. Although single-opponent cells have a clear visual function for processing the color region and luminance contrast, their role in CC is not considered important because it is usually held that their color response to stimuli is not illuminant-independent \cite{conway2010advances,gegenfurtner2003cortical,shapley2011color}. 
%Concretely, two types of single-opponent neurons with different RFs were described, in which Type II has only the color-opponent RF (RF) in the center, which mainly respond to a uniform color area, while type I has a center-surround RF with color-opponency \cite{livingstone1984anatomy}. This type of center-surround color-opponent RF structure can not only respond to the uniform color regions, but also respond well to the brightness contrast of images such as the luminance-defined edges \cite{shapley2011color}. Although single-opponent cells have clearly visual function for processing color region and luminance contrast, their role with CC is irrelevant because it is usually thought that their color response to stimuli is not illuminant-independent \cite{conway2010advances,gegenfurtner2003cortical,shapley2011color}. 

Earlier studies have indicated that the responses of the color-selective neurons in V1 are wavelength-dependent instead of wavelength-differencing dependent as proposed in a famous study \cite{zeki1983colourresponses}. In contrast, the results show that the color-selective neurons in V4 are more wavelength-differencing dependent and hence may contribute to CC and the color contrast phenomenon of the HVS. However, none of the previous results have clearly identified the receptive field (RF) of color-selective neurons in detail. According to one hypothesis, there should be a kind of DO neuron with the spatial center-surround opponent RF and color-opponency \cite{michael1978color}. Its function may be crucial for CC \cite{conway2010advances,gegenfurtner2003cortical,shapley2011color}.

%Although there are also many color-selective neurons recorded in V1, 
%Earlier studies pronely indicated that the responses of color-selective neurons in V1 are wavelength dependent instead of wavelength-differencing dependent as proposed in a famous study \cite{zeki1983colourresponses}. In contrast, their results show that color-selective neurons recorded in V4 are more wavelength-differencing dependent and hence may contribute to CC and color contrast phenomena of HVS. However, none of the previous results clearly parsed out the RF of color-selective neurons in details but proposed a hypothesis that there should be a kind of double-opponent neuron with the spatial center-surround opponent RF and the color-opponency \cite{michael1978color}. Its function may be very important for CC \cite{conway2010advances,gegenfurtner2003cortical,shapley2011color}.  
Electrophysiological experiments have found that there are also DO neurons in the V1 of rhesus monkeys \cite{johnson2001spatial,de2021spatial,conway2001spatial}, and they have both the center-surround spatial-opponent and color-opponent RFs, which respond to specific local color patterns. Due to their spatial-opponent and band-pass filtering properties, these DO neurons can respond well with orientation selectivity \cite{de2021spatial,johnson2001spatial}. The typical spatial RFs of DO neurons can be fitted well with a two-dimensional (2D) Gabor function or a non-concentric version of the Difference of Gaussian function (DoG) with a crescent-shaped surround \cite{de2021spatial}.

%Until now, electrophysiological experiments have really found that there are also the double-opponent neurons in the V1 of rhesus monkeys \cite{johnson2001spatial,de2021spatial,conway2001spatial}, %. Experimental results clearly show that the 
%where double-opponent neurons indeed have both center-surround spatial-opponent and color-opponent RFs, and respond to the local color pattern. At the same time, due to their spatial-opponent and band-pass filtering properties, these double-opponent neurons can well respond to the orientation \cite{de2021spatial,johnson2001spatial}. %orientational color gratings \cite{de2021spatial,johnson2001spatial}. 
%The double-opponent neurons may be further divided into the center-surround double-opponent neurons and the orientation-selective double-opponent neurons according to the spatial structure of their receptive fields \cite{gegenfurtner2003cortical}. Computational modeling study \cite{yang2015boundary} shows that the double-opponent neurons with orientation selectivity are able to simultaneously detect boundaries defined by brightness and color in images. 
%The typical spatial RF of double-opponent neurons can be well fitted with a Gabor function or a non-concentric version of Difference of Gaussian function (DoG) with a crescent-shaped surround \cite{de2021spatial}. 

The debate on whether DO neurons in V1 exist was finally resolved, and their color RFs and spatial integration properties in detail were clearly identified very recently \cite{conway2001spatial,johnson2001spatial}. However, the role of DO neurons in image processing and visual perception is still not clear \cite{de2021spatial}. In this work, we focus on V1 to identify its possible CC mechanisms because many electrophysiological experiments have pointed out the existence of color-selective neurons in V1 and indicated the existence of DO neurons that are hypothetically treated as the perfect computation unit for archiving CC owing to their special RFs \cite{gegenfurtner2003cortical,shapley2011color}. However, these theoretical descriptions usually either just propose that V1 contributes CC by discounting the illuminant \cite{conway2010advances,gegenfurtner2003cortical} or by performing illumination-independent responses \cite{zeki1983colourresponses}, and they do not provide any reasonable algorithms describing how V1 DO neurons achieve CC. Our hypothesis is different from the theoretical descriptions mentioned above. %We propose that some of the V1 double-opponent neurons encode the illuminant or perform the illumination-dependent responses for achieving CC. %Concretely, we propose that the function of V1 double-opponent neurons is not discounting the illuminant or performing illumination-independent responses. 
In contrast, we propose that some of the V1 DO neurons serve to encode the illuminant or perform the illumination-dependent responses for achieving CC. 

Our inference is based on two previous studies. One is a neuroimaging result relying on a psychophysics experiment, which indicates that V1 largely tends to encode the illuminant of the scene than the other advanced visual areas such as V2, V3, and V4 in the HVS \cite{bannert2017invariance}. {\color{black}In addition, fMRI scanning combined with a model-based decoding approach shows that the early visual areas, such as V1 and V2, are modulated by the chromatic light stimuli. However, the higher visual areas, such as V4 and VO1, correspond strongly to the representation of perceived color \cite{kim2020neural}.} Another evidence is our previous modeling work related to DO neurons, which helps discover that the image processing responses of DO neurons to a color-biased image effectively contain the information of the illuminant \cite{gao2015color}. 
Based on these limited evidence, we address two questions in this work: (1) whether a simple non-linear V1 neural model can learn to encode the illuminant; and (2) what type of RF can be learned by the V1 neural model if it can predict the illuminant with good accuracy?

The electrophysiologically based V1 neural model we build mainly consists of a linear convolution layer, a non-linear divisive normalization layer, and a decoding layer with two fully connected layers. We train the model to predict the illuminant (an image-computable quantity) from natural images captured by real cameras using the light source color estimation as the optimization objective. This model structure proves capable of capturing the V1 neural responses to natural images and obtaining better performance in predicting neural spike responses than the classical subunit model \cite{burg2021learning}. By this modeling framework, we can understand how each component of the V1 neural model contributes to illuminant prediction. We test our hypothesis that DO cells in V1 serve to predict rather than discount the illuminant in achieving CC. In addition, we also devise a baseline task of discounting the illuminant to illustrate that estimating the illuminant-independent invariance such as the mean of edge responses cannot guide the V1 model to learn the meaningful DO RF.

%Different from the previous work \cite{burg2021learning} that lets the model learn to predict spike responses in V1 to natural images recorded by electrophysiology experiments, we train the similar model to predict the illuminant (an image-computable quantity) from natural images captured by the real cameras using the light source color estimation as an optimization objective. One of novelties of our paper is that previous work on RF analysis was based on high-level constraint such as object classification and recognition that seems to be far away from CC \cite{yamins2016using}, but our approach is for the first time based on the constraint of light source color estimation. %, which may be more conform an image processing function of early visual cortex such as V1. 
%By this modeling framework, we would see how each component of V1 neural model contributes to illuminant prediction (IP) and test our hypothesis that double-opponent cells in V1 serve to predict rather than discount illuminant. In addition, we also devised a baseline task of discounting the illuminant to illustrate that estimate the illuminant-independent invariance such as the mean of edge responses can not guide the V1	model to learn the meaningful double-opponent RF.

\section{Results}\label{sec3}
\subsection*{Illuminant estimation accuracy of four V1 models} 
%In this section, we first show the comparison of illuminant prediction accuracy among four V1 models including standard subunit model, the DN model with and without surround, and CNN. Then, we show the visualization and analysis of trained receptive fields of each model. We use the singular value decomposition (SVD) to decompose the trained receptive fields into the color weight and spatial weight \cite{horwitz2005paucity}, and qualitatively compare the trained receptive fields by models with the actually measured receptive fields on Monkey's V1 \cite{de2021spatial}. %(1), comparison of light source estimation accuracy (2), comparison of light source estimation performance with and without DN structure, CNN3.
\begin{figure*}[!h]
	\includegraphics[angle=0,width=1\textwidth]{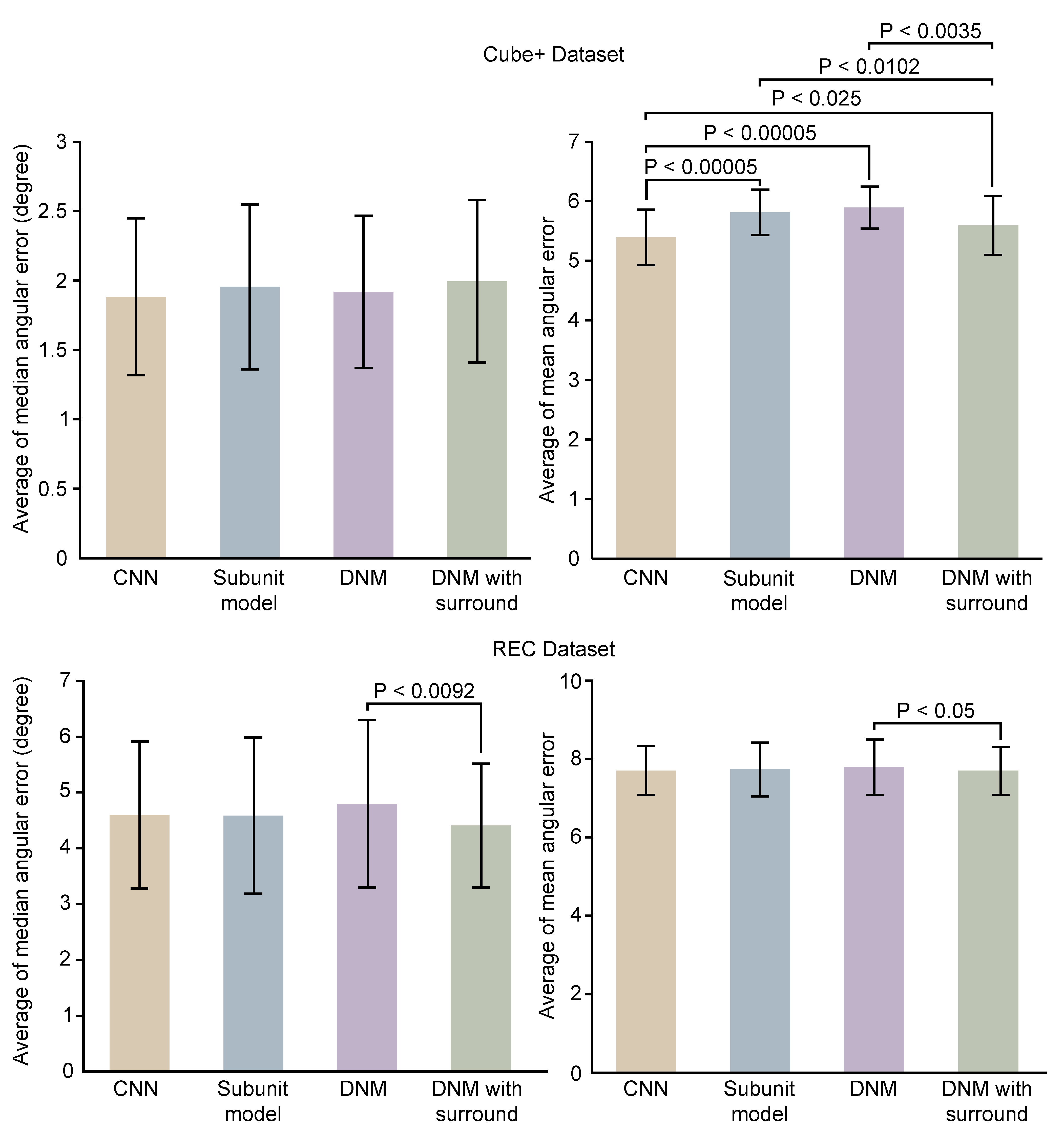}
	\caption{Comparison of four simulated V1 neural models on predicting illuminant. The left is the average of the median angular error of 60 results on the whole Cube+ dataset. The right is the average of the mean angular error of 60 results on the whole Cube+ dataset. Each error bar indicates the standard error of the mean. The p values labelled above the bar are obtained using the pairwise Wilcoxon signed rank test on 60 results of the prediction accuracy on three validation folds. Similar interpretation applies to the results on the REC dataset in the second row, in which the statistics are obtained on 120 results of the prediction accuracy on three validation folds.
	}
	\label{Figure2}
\end{figure*}

Figure. \ref{Figure2} shows the performance of illuminant prediction (IP) of four V1 neural models on two benchmark datasets (see Methods for details). In terms of the median angular error, four V1 neural models achieve good IP performance on both the Cube+ and REC datasets tested here. For example, the four V1 models from the CNN to the DNM with surround obtain an average median angular error of 1.8826, 1.9545, 1.9187, and 1.9943 on the Cube+ dataset. Although directly comparing these solutions is not applicable, two state-of-the-art deep learning approaches report their best single median angular errors as 1.32 \cite{kovsvcevic2020guiding} and 0.93 \cite{hu2017fc}. The proposed V1 approach is a quite simple biological model despite its average performance being not as good as the single best indicator of engineering algorithms. Although there is no significant difference among these models in terms of the median angular error except for the DNM and the DNM with surround on the REC dataset, this result clearly shows that the quite simple V1 neural model can be trained to predict the illuminant from the input image. Generally, the simpler subunit model, DNM, and DNM with surround consistently achieve good performance on the two benchmark datasets compared to the deep CNN model with a more complex structure and more parameters.

In terms of the mean angular error, both the CNN and the DNM with surround perform clearly better than the other two models, namely the subunit model and the DNM, which demonstrates that the CNN and the DNM with surround are more robust than the subunit model and the DNM. Notably, the DNM with surround outperforms the subunit model and the DNM by only adding $7\!\times\!7$ surround-based divisive normalization, showing the importance of surround-based divisive normalization as an additional computational mechanism to obtain robust IP from the input images.
%Furthermore, in terms of mean angular error, both models of CNN and DNM with surround performs clearly better than other two models including subunit model and DNM, which demonstrates that CNN and DNM with surround are more robust than the subunit model and DNM. Notably, DNM with surround outperforms subunit model and DNM by only adding $7\!\times\!7$ surround based divisive normalization, showing the importance of surround based divisive normalization as the additional computational mechanism to obtain the robust IP from the input image.
\begin{figure*}[!h]
	\includegraphics[angle=0,width=1\textwidth]{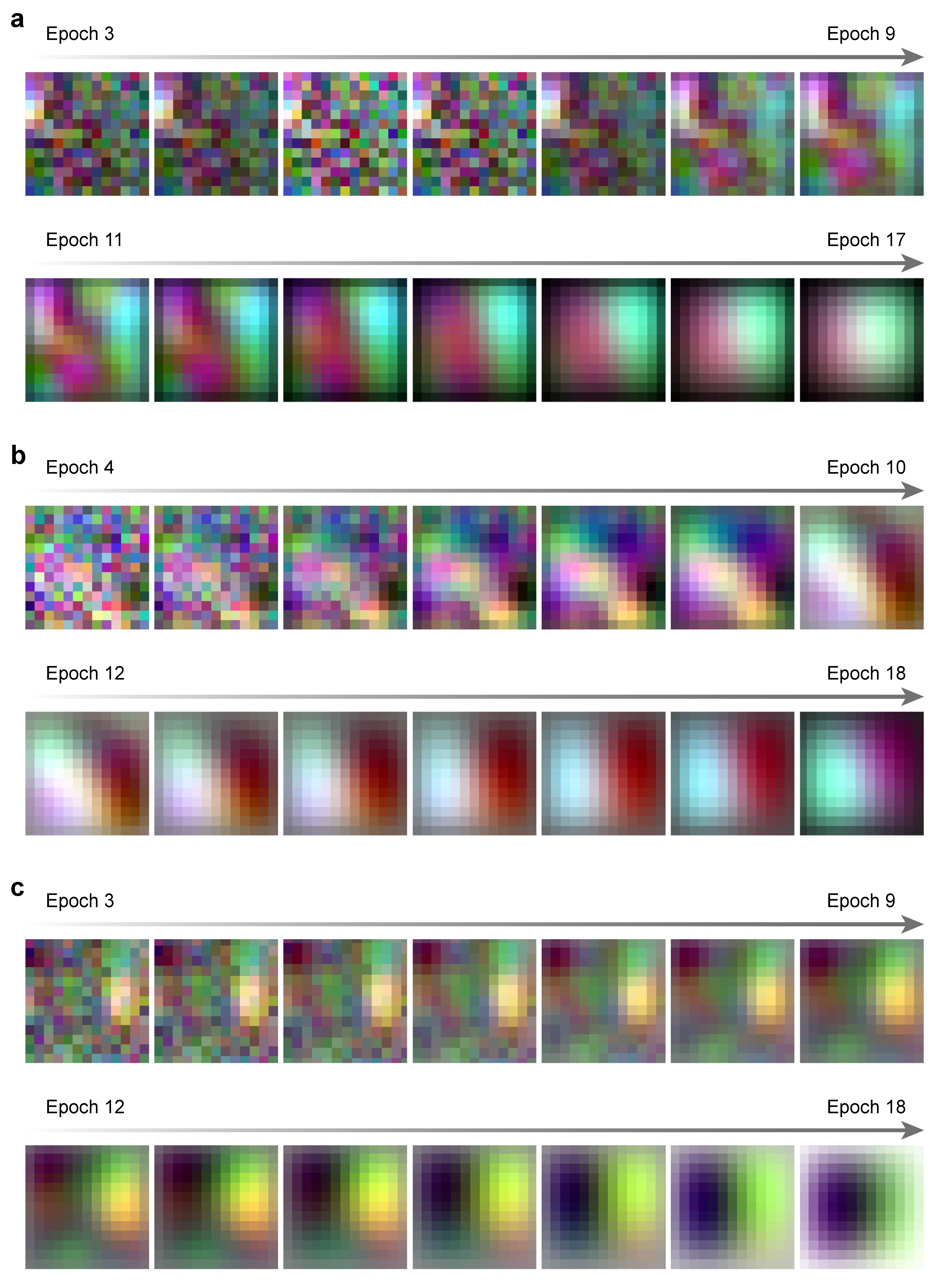}
	\caption{%{\bf Bold the figure title.}
		Trained convolution kernels during the iteration of the epochs. The random weights of the kernel gradually converge to a color-opponency subunit structure (a) and (b) red-green color-opponency RF, and (c) blue-yellow color-opponency RF.}
	\label{Figure3}
\end{figure*}
\subsection*{Models learn various RFs physiologically observed in V1 when predicting illuminant} 
%A key question is that what’s the color and spatial receptive fields of V1 neuron looks like? Specially, we want to see whether the trained model can predict the color and spatial receptive fields observed in V1, which have been recently parsed out using the physiological experiments. Another question is whether the four V1 neural models simulated with quite different model structures (e.g., CNN and DNM) could learn different color and spatial receptive fields under a single constraint that is illuminant estimation? 

\begin{figure*}[!h]
	\includegraphics[angle=0,width=1\textwidth]{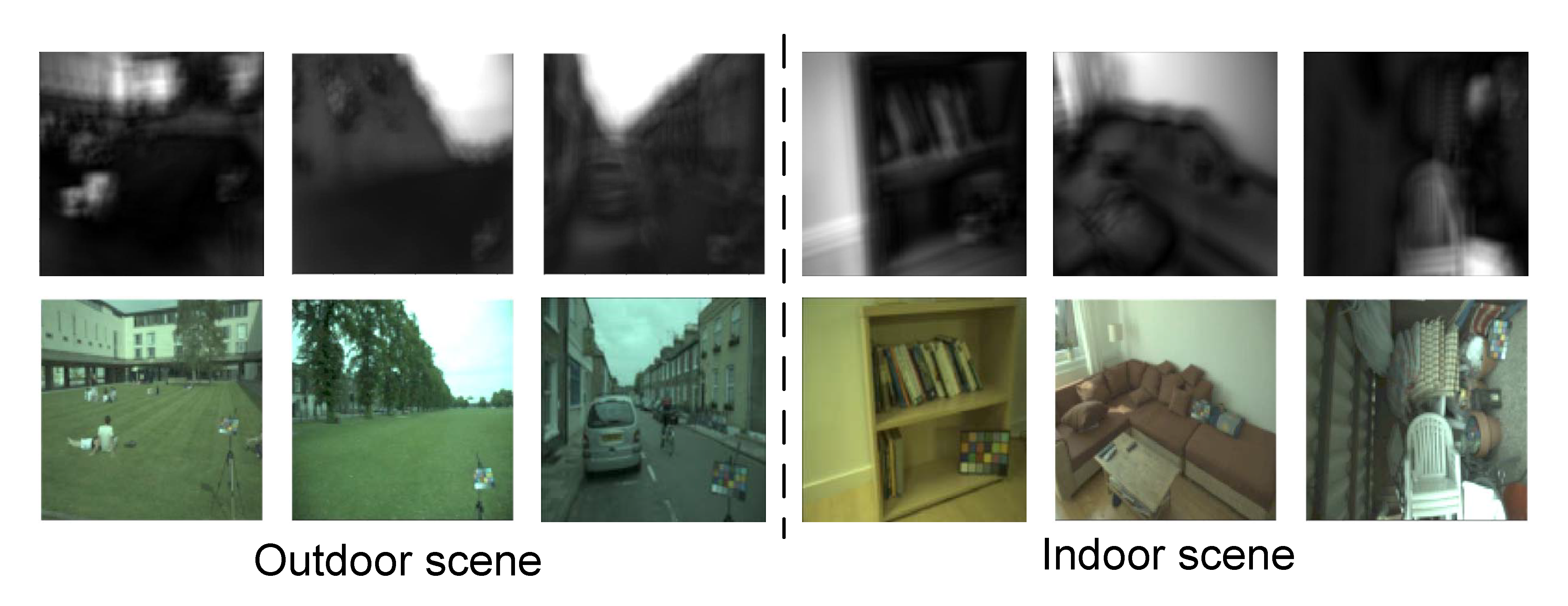}
	\caption{%{\bf Bold the figure title.}
		Responses of DN for several examples of indoor and outdoor scenes. % The images are Gamma corrected to increase the luminance and contrast for better visualization. %We can observe that responses of DN primarily focus on the regions such as the sky that maximally encode the illuminant information and significantly ignore other regions. 
	}
	\label{Figure11}
\end{figure*}
\begin{figure*}[!h]
	\includegraphics[angle=0,width=1\textwidth]{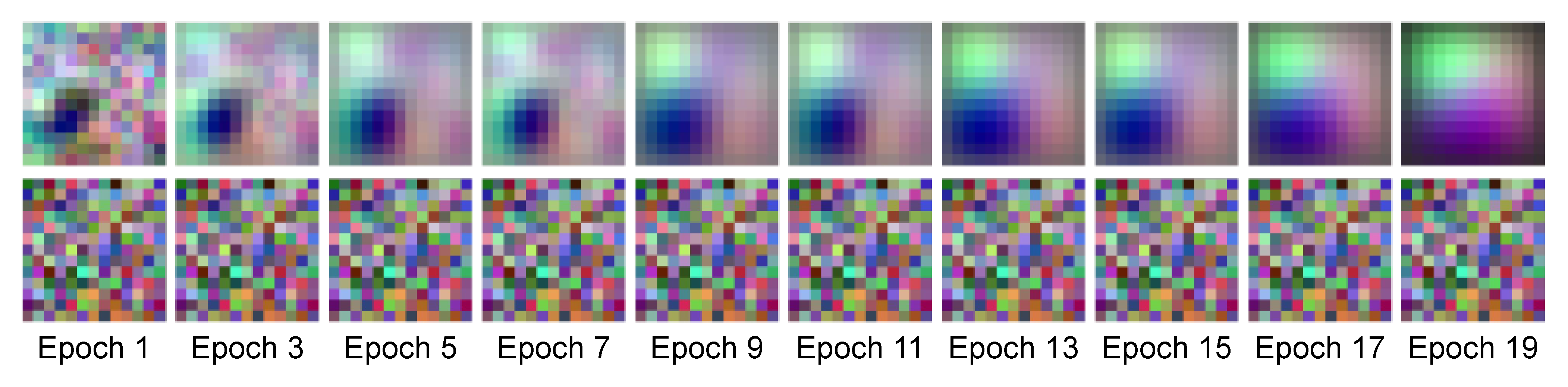}
	\caption{%{\bf Bold the figure title.}
	Trained convolution kernels during the iteration of 20 epochs under conditions of predicting the illuminant (the first row) and discounting the illuminant (the second row). Images are displayed for every other epoch.}
	\label{Figure5}
\end{figure*}
Figure. \ref{Figure3} visualizes the trained convolution kernels (i.e., the learned RFs) under different training epochs. Each epoch represents a complete training of the model on the training dataset. We can see that the RFs gradually converge to a stable state with a clear spatial structure and color subunits from the initial state with random noise. The learned RFs are quite similar to the RFs of the DO neurons recorded in V1 \cite{de2021spatial}. For example, the first row of Figure. \ref{Figure3} shows a learned RF with red-green color opponency, which is similar to the spike-triggered average (STA) results of Fig. 3C in De and Horwitz \cite{de2021spatial}. Similarly, the third row of Figure. \ref{Figure3} shows a learned RF with blue-yellow color opponency, which is similar to the STA results of Fig. 3E in De and Horwitz \cite{de2021spatial}. Both the STA results of Fig. 3C and Fig. 3E in De and Horwitz \cite{de2021spatial} are adapted as Figure. \ref{Figure14} for comparison in the supplementary information (SI).
%Both STA results of Figure. 3C and Figure. 3E in \cite{de2021spatial} are adapted as Figure. \ref{Figure14} for comparison in supplementary information.

We visualize the responses of each neuron after divisive normalization (DN) processing (see Methods) in Figure. \ref{Figure11}, which shows that the responses of the DN primarily focus on specific regions (e.g., the sky in the image of the second column) that maximally encode the illuminant information. In other words, these results clearly show that the neurons in V1 are computationally attending to the illuminant. For comparison, the V1 model trained to discount the illuminant failed to learn the effective RF structure as shown in Figure \ref{Figure5} compared with the model trained to predict the illuminant.

\begin{figure*}[!h]
	\includegraphics[angle=0,width=1\textwidth]{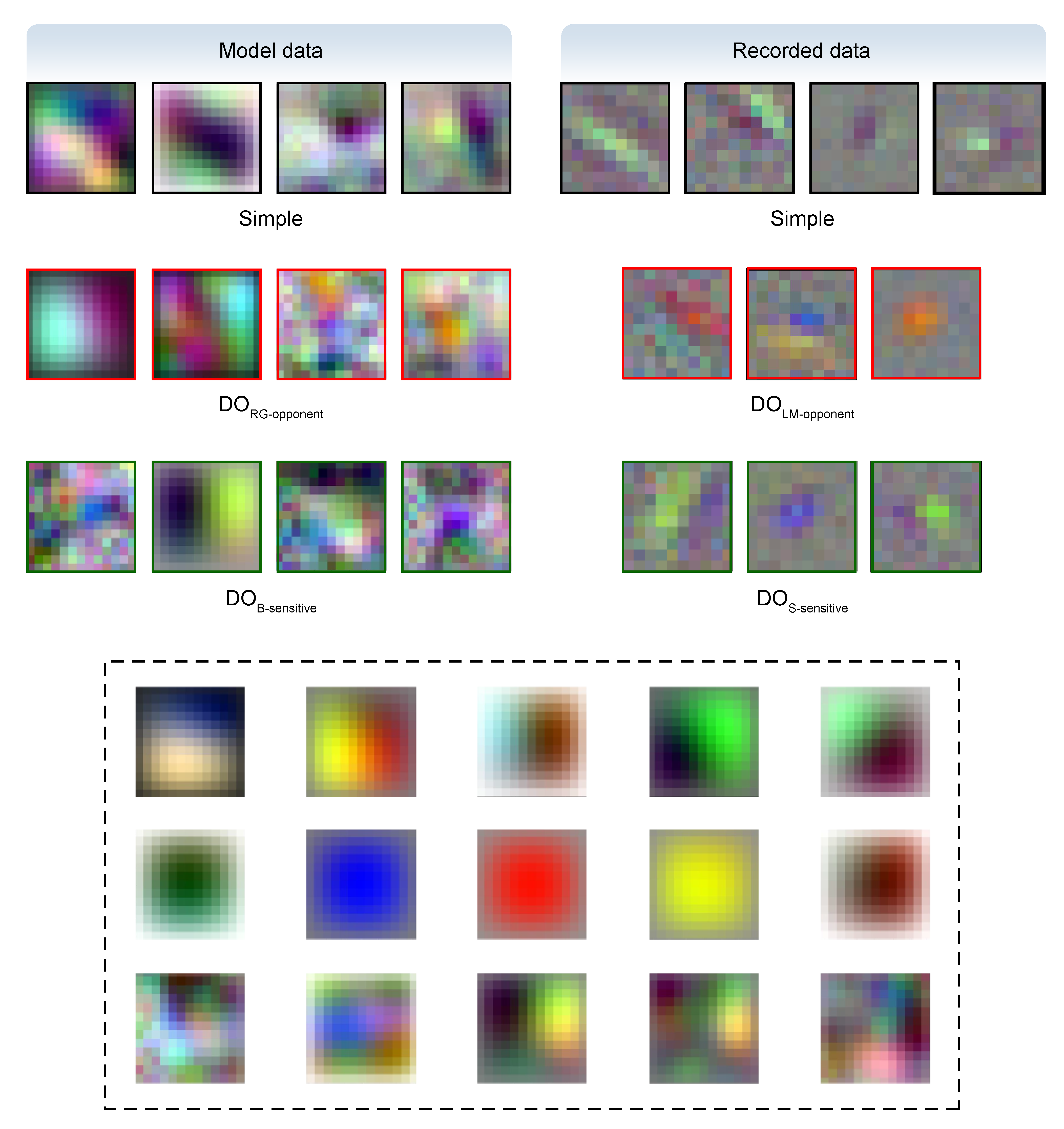}
	\caption{%{\bf Bold the figure title.}
		Samples of RFs of V1 neural models learned from the IP task. The learned RFs can be classified into several types, such as localized and oriented, which are quite similar to the simple, $DO_{LM-opponent}$, and $DO_{S-sensitive}$ RFs in V1 neurons recorded using the spike-triggered average (STA) method \cite{de2021spatial}. Some RFs in the dotted line box show clear color-opponency structure (see the descriptions in the main text).
		The recorded data are adapted from Figure 3 and Figure 4 in De and Horwitz \cite{de2021spatial}.
		%Some RFs labeled with the dotted line box show the clear color-opponency structure in the first row images. Some shows the typical Gaussian-shape based color RF in the second row images. There are also some RFs show several color subunits within RF in the last row images.
	}
	\label{Figure13}
\end{figure*}
Figure. \ref{Figure13} shows other RFs learned by the V1 neural models, which can be classified into several types. Some RFs show the typical localized and oriented structures (e.g., model data) that are quite similar to the recorded RFs in V1 neurons using the spike-triggered average method (e.g., recorded data). The model data can be roughly classified as three types of cells, namely simple cells, $DO_{\!RG-opponent}$ cells, and $DO_{\!B-sensitive}$ cells, according to the criterion of the recorded V1 data \cite{de2021spatial}. %, where the corresponding neurons are classified into simple cells, double-opponent cells with L-M opponency, double-opponent cells with S-sensitive \cite{de2021spatial}.

In addition, some RFs in the dotted line box of Figure. \ref{Figure13} show clear color-opponency structure (e.g., the first image in the first row shows the blue-yellow color opponency RF structure, and the fifth image in the first row shows the red-green color opponency RF structure). Other RFs show the typical Gaussian shape-based structure with color selectiveness (e.g., the images in the second row). There are also some RFs that show several color subunits within the RF (e.g., the images in the third row).
%In the next, we will adopt the standard mathematical approach to further analyze the learned RFs of V1 neural models.
\subsection*{Quantitative analysis of the color weight function and spatial weight function using SVD} 
%In order to quantitatively describe the learned RFs, 

\begin{figure*}[!h]
	\includegraphics[angle=0,width=1\textwidth]{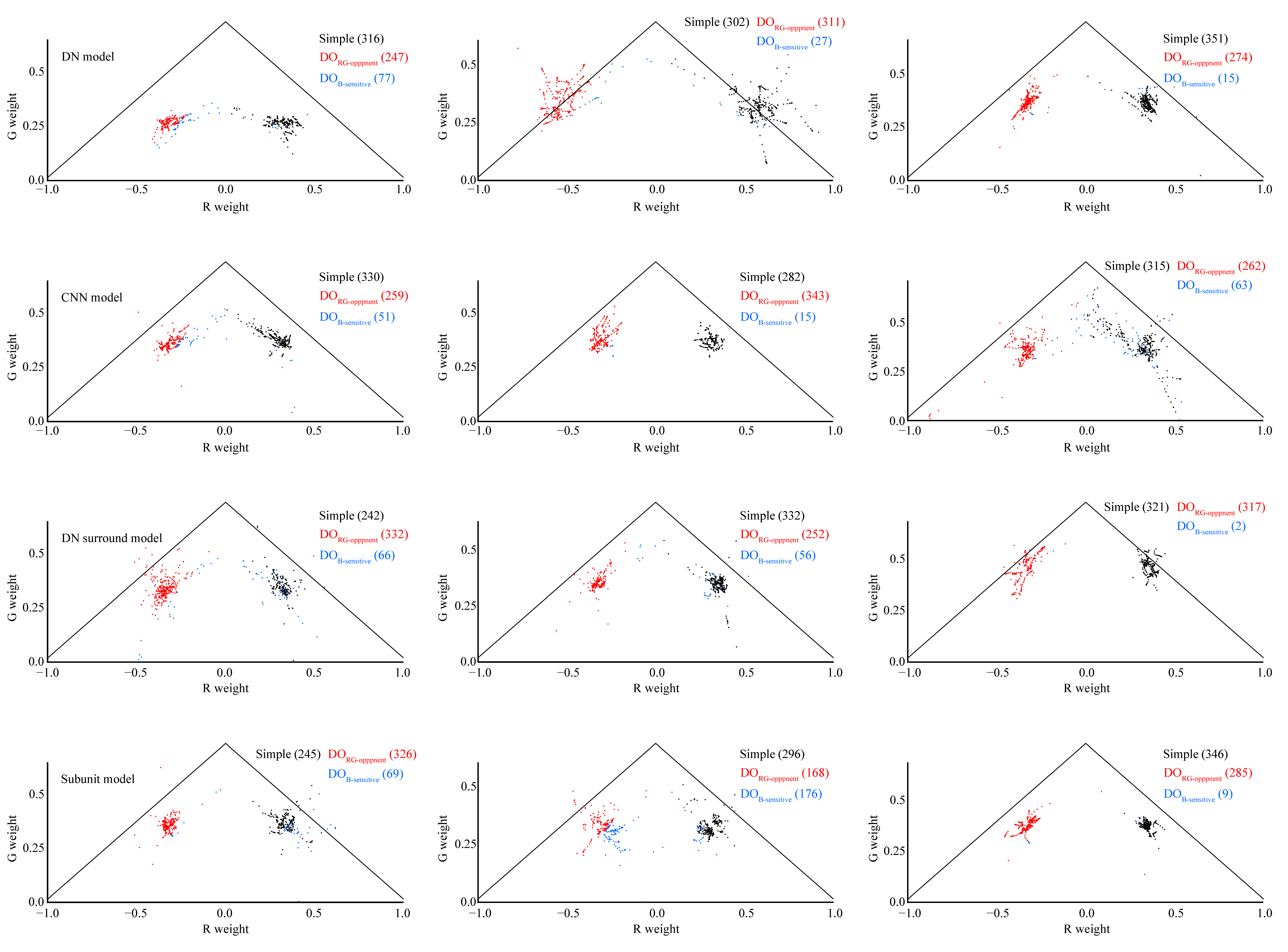}
	\caption{%{\bf Bold the figure title.}
		Cell category distributions of trained RFs based on their color weights on the Cube+ dataset \cite{banic2017unsupervised}. From top to bottom: the DN model, the CNN model, the DN model with surround, and the subunit model. From left to right: fold 1, fold 2, and fold 3, where each fold contains 20 results and each epoch includes 32 trained kernels (i.e., the RF of the model neuron). Therefore, each sub-figure consists of 640 model neurons. The RFs are classified as simple cells (labeled with black points), DO cells with red-green color opponency (labeled with red points), and DO cells with blue-sensitive color opponency (labeled with blue points). The criteria of the normalized weight classification are roughly based on De and Horwitz \cite{de2021spatial}. See Methods for details.}
	\label{Figure16}
\end{figure*}

We adopted the modified singular value decomposition technique (SVD) as used in \cite{de2021spatial,horwitz2005paucity} to decompose the RFs into color weights and spatial weights. Figure. \ref{Figure16} shows all results of the classified normalized color weight for each model neuron's RF for four V1 neural models after training on the Cube+ dataset for IP. Similar results for the REC dataset are shown in SI Figure. \ref{Figure5_RCC}.  
We find that the classified normalized color weight distributions are quite similar to the physiological results of \cite{de2021spatial}. We observe a similar trend as the physiological report, that is, most recorded neurons are simple neurons because both the R and G weights have the same signs. The second-highest type of recorded neurons is DO neurons with red-green color opponency because their R and G weights have opposite signs. The last type of neurons, with a relatively low number, is the DO neurons with blue-yellow color opponency because their B weights have signs opposite to that of the R and G color weights, and they account for a large proportion of the total variance. This analysis shows that the four V1 neural models can learn meaningful RFs only under the constraint of IP. Some learned V1 neurons posses DO RFs, and hence, we can conclude that one of the functions of V1 neurons (e.g., at least some DO neurons shown in Figure \ref{Figure16}) is to encode the illuminant of a scene. %We will further prove this inference in the next section.

Generally, there are two differences between our results and the physiologically recorded results in V1. First, the normalized weight distributions in our results are visually more concentrated. As a comparison, the normalized weight distributions in Figure 2 of \cite{de2021spatial} are visually more spread. Second, De and Horwitz included the normalized weights of the neurons that can not be classified into simple neurons or DO neurons according to their weight criteria. In contrast, we have no this issue, and all our normalized weights can be classified into three categories according to our slightly revised criteria (e.g., we classify neurons as DO blue-sensitive neurons if only their blue channel weights are opposite to that of the other two color channels). Hence, we may have observed more blue-sensitive neurons learned by our models than that in the results in \cite{de2021spatial} (e.g., see some examples of visualization results in Figure. \ref{Figure13}, where the RFs of some blue-sensitive neurons are plotted).

\begin{figure*}[!h]
	\includegraphics[angle=0,width=1\textwidth]{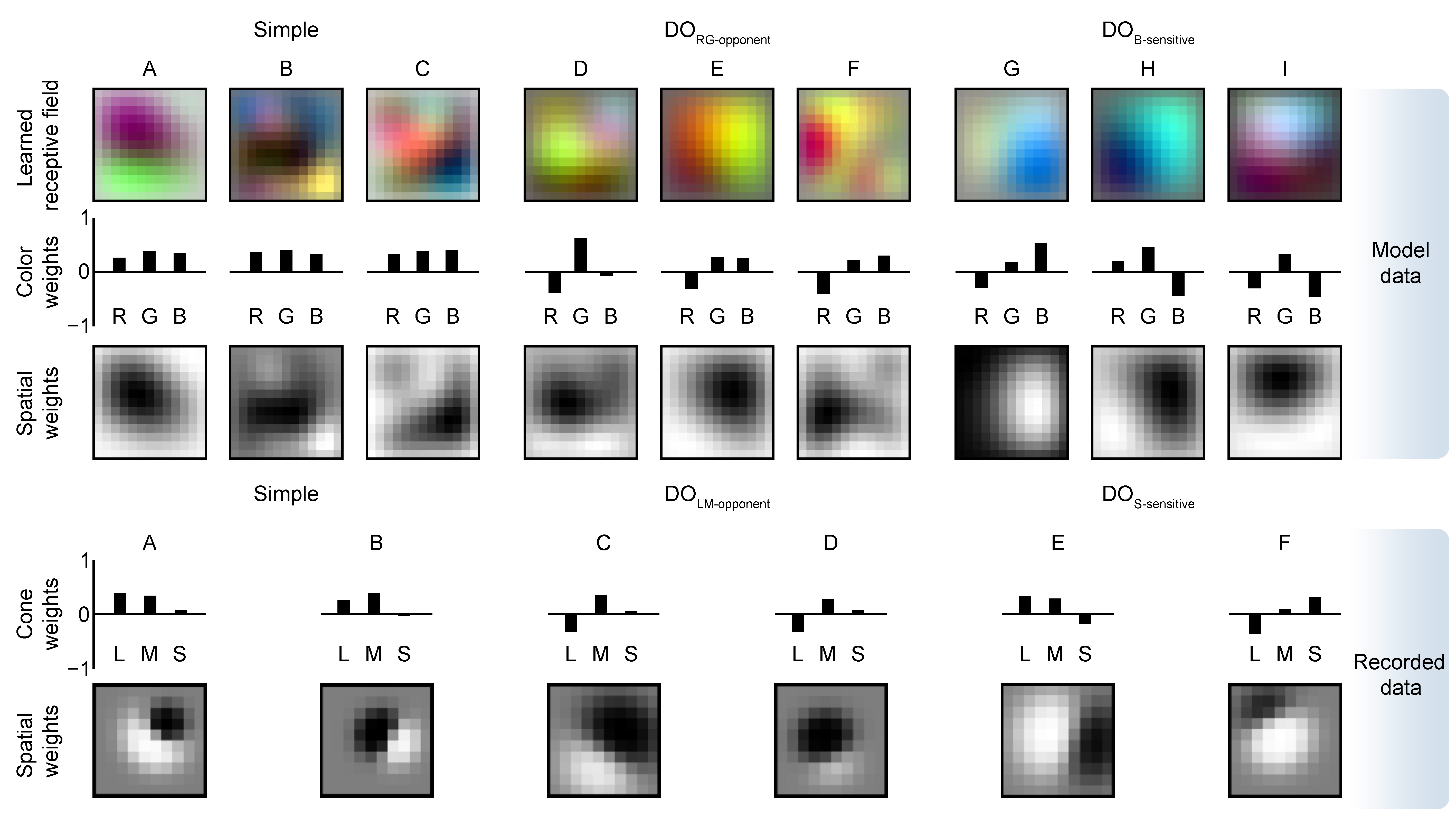}
	\caption{%{\bf Bold the figure title.}
		The first row shows the decomposed results of nine example neurons using singular value decomposition (SVD). Each model neuron's RF is decomposed into the color and spatial weights. The classification criterion is roughly according to De and Horwitz \cite{de2021spatial}. The second row shows the decomposed results of six neurons recorded in V1 using SVD. The spatial weights are obtained using the non-concentric DoG fitting. The recorded data are adapted from Figure 3 and Figure 7A of De and Horwitz \cite{de2021spatial}.}
	\label{Figure12}
\end{figure*}

Figure. \ref{Figure12} shows some examples of the learned color weights and spatial weights using SVD, which are quite similar to the physiological ones recorded by De and Horwitz \cite{de2021spatial} (model data vs. recorded data). However, the decomposed spatial weight distributions based on the model RF have a clearer spatial structures than previously recorded and fitted using the DoG or a Gabor function. Note that we did not adopt any functions to fit the trained RF data.

The computational model results (e.g., the spatial weights in Figure. \ref{Figure12}) support that the spatial RF of the V1 DO neuron is closer to a Gaussian function surrounded by a semicircle (e.g., the third row of the model data), which is roughly consistent with the conclusion that a circular center with a crescent around fits the data as well as the Gabor function \cite{de2021spatial} (e.g., the spatial weights in the second row of the recorded data).

 %In our analysis, qualitatively, the spatial RF is closer to a Gaussian or a Gaussian surrounded by a semicircle, rather than a center-periphery structure, which supports the results of the electrophysiological experimental analysis. %(Re-unify the visualization of the weights, there are original images and images separated from SVD, and analyze the visualization of the weights of R, G, and B at the same time, and how the weights are normalized)

%Therefore, it shows that V1 is indeed estimating the light source color, and V1 double-opponent neurons specifically implement light source color estimation. . . ? On the contrary, if four different models are learned differently, it does not mean that V1 is estimating the color of the light source.
\subsection*{DO cells perform more robustly than simple cells in V1 for IP} 
\begin{figure*}[!h]
	\includegraphics[angle=0,width=1\textwidth]{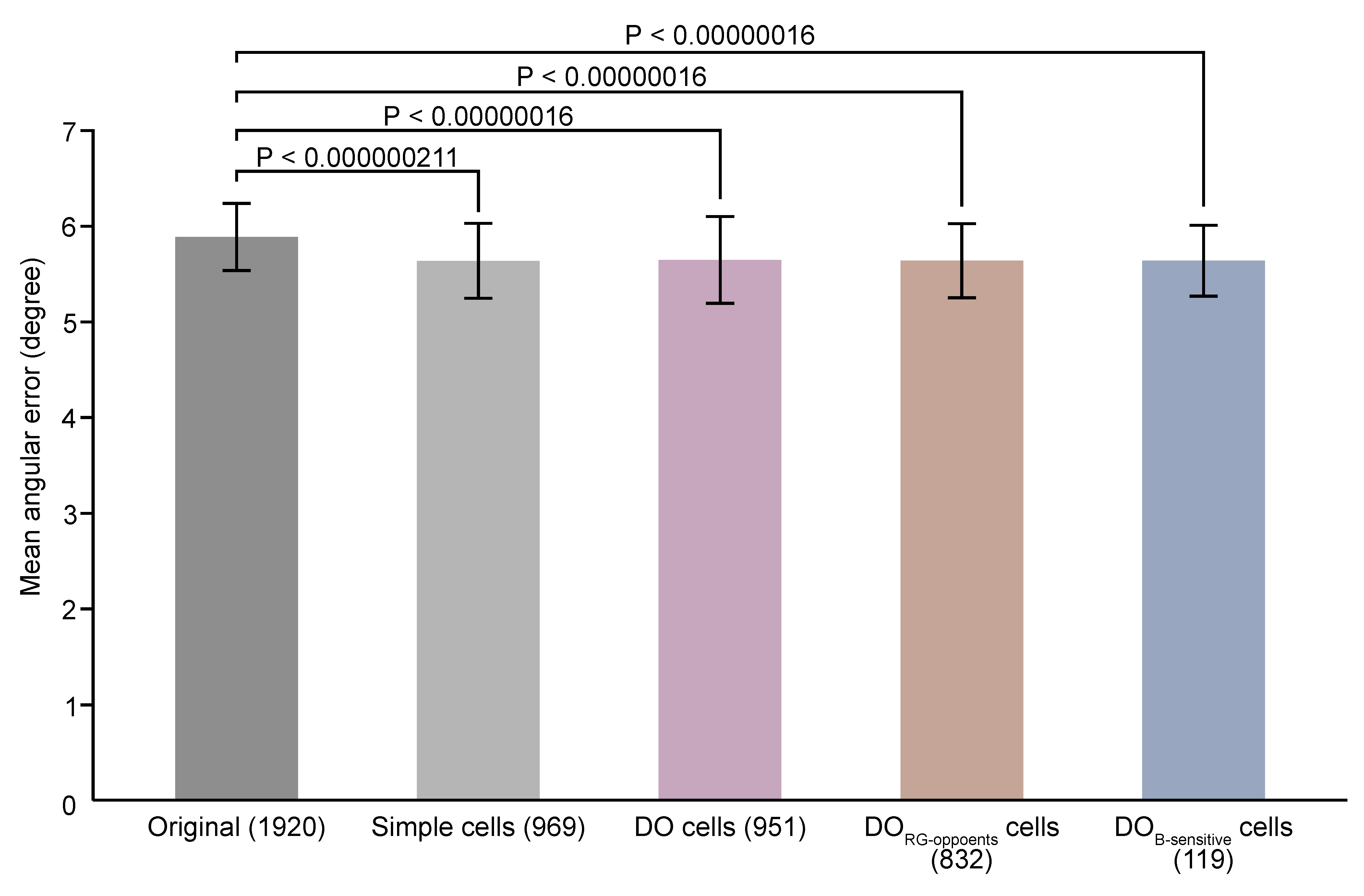}
	\caption{%{\bf Bold the figure title.}
		Results of the DNM using the randomized weights of the DO cells and simple cells on three folds of the test dataset (the Cube+ dataset). The bar shows the mean angular error of 60 results on the whole dataset. Each error bar indicates the standard error of the mean. The p values labelled above the bar are calculated using the pairwise Wilcoxon signed rank test on 60 results of the prediction accuracy on three validation folds. Note that only the p values smaller than 0.05 are labeled. The five bars represent different categories of V1 cells, and the number in brackets indicates the number of the corresponding neuron category used for the prediction.}
	\label{Figure7}
\end{figure*}
In order to ascertain which types of V1 neurons exhibit greater performance in IP, we modify a V1 model by randomly assigning the weights of either the simple cells or the DO cells. Figure. \ref{Figure7} shows the experimental results of V1 model using the randomized weights of the DO cells and simple cells on three folds of the test dataset (see Methods). The randomized DO cells, namely the DO cells with red-green color opponency and the DO cells with blue color-sensitive, account for 49.53\% of the total cells compared to the simple cells accounting for 50.47\% of the total cells  (e.g., 951 DO cells vs. 969 simple cells in three folds). 
%We can observe that the V1 model only using the weights of the DO cells or only using the weights of simple cells performs significantly better than the original V1 model comprising 1920 cell weights in its entirety.
We notice that the V1 model, when utilizing only the weights of the DO cells or solely those of the simple cells, demonstrates notably superior performance compared to the original V1 model containing all 1920 cell weights.

No significant statistical difference is observed in the performance of the V1 model with only DO cells compared to that of the model with only simple cells. Interestingly, no significant statistical difference is observed in the performance of the V1 model with only $DO_{RG-opponent}$ cells and that of the V1 model with only $DO_{B-sensitive}$ cells compared to the performance of the V1 model with only simple cells. In other words, the V1 model with DO cells shows similar performance as the simple cell-based V1 model while using very few neurons. For example, the $DO_{B-sensitive}$-based V1 model contains only 119 neurons, which account for approximately only 10\% of the total number of neurons of the simple cell-based V1 model. These results quantitatively demonstrate that the DO cell-based V1 model performs more robustly in IP than the simple cell-based V1 model.

\section{Discussion}\label{sec4}
Our computational results indicate that the four V1 models with quite different structures (e.g., some consisting of only the linear unit, while others consisting of both the linear and nonlinear units), constrained by an IP task, can learn the receptive field structures similar to those of the V1 neurons, including DO and simple cells. DO cells have long been considered to be very important for achieving visual color constancy \cite{gegenfurtner2003cortical,shapley2011color}, but how DO cells achieve color constancy is unclear. In this paper, we explicitly point out through a computational study that some DO cells can encode the illuminant color of a scene. Our approach is the first work based on a low-level constraint that may be more consistent with a visual perception function of the early visual cortex, such as V1. We discuss the computational study results from a mathematical perspective of reducing redundancy (see mathematical derivation in the SI) to explain why DO cells or other cells in V1 may encode rather than discount the illuminant.

%In summary, we observed that the four V1 models with quite different structures (e.g., some just contain the linear unit, some contain both the linear and the nonlinear units) can learn the RF structure similar to V1 neurons including double-opponent neurons and simple cells only under the constraint of light source color estimation. 

From the view of efficient coding, the spatial-opponency of DO cells is not to eliminate the color of the light source but to constitute a low-pass or band-pass filter with the center-surround RF, which can effectively extract low-frequency or intermediate-frequency light source color information \cite{gao2015color} for the subsequent DN stage for reducing redundancy. The function of color-opponency of the DO cells is also to eliminate the redundancy due to the large overlap of light-sensitive responses by the L, M, and S channels to achieve efficient coding of natural images \cite{lee2002color}. The reduction of redundancy achieved through color-opponency can be likened to optimizing the responses of the L, M, and S channels towards an impulse function, rather than a broader overlapping Gaussian function.
%This elimination of redundancy through color-opponency is somewhat similar to letting the responses of the L, M, and S channels be based on an impulse function as much as possible rather than an overlapping wider Gaussian function. 

{\color{black}Based on our findings, we conclude that V1 neurons, including simple cells, have the ability to encode the color of the light source. However, DO cells are particularly robust and efficient in encoding the color of the light source due to their specialized color-opponency. This color-opponency allows for independent encoding of the L, M, and S signals, ensuring that the encoded signals meet the requirements for diagonal or von Kries transformations, which are essential for independent color channel correction \cite{funt2000diagonal}. The impulse function-style response of DO cells paves the way for the subsequent step of extracting an accurate light source color based on the spatial-opponency of these cells. The combined output of this process, along with the final DN-based V1 neuron, effectively eliminates redundancy in light source color and other global information, resulting in an invariant response. This response represents the local color characteristics of the true image, as assumed by the Gaussian scale mixture model \cite{coen2015flexible}. In summary, DO cells, as well as other cells in V1, encode instead of disregarding the illuminant, which is a valuable statistical information about the surrounding environment. This information can be further utilized in computations, such as DN, to accurately infer the true invariant characteristics of the image or object, such as its intrinsic color. This efficient coding approach allows for effective utilization of available information.}

%Therefore, our inference is that V1 neurons, including simple cells, can encode the color of the light source, but the reason that the double-opponent cells encoding the color of the light source are more robust and efficient is their special color-opponency, which realizes independent encoding of the L, M, and S signals so that the encoded signals satisfy the preconditions of diagonal transformation or the von Kries transformation for independent color channel correction \cite{funt2000diagonal}. The impulse function-style response paves the way for the second step to extract a valid light source color based on the spatial-opponency of DO cells, both of which together serve the output of the final DN-based V1 neuron, eliminating light source color and perhaps other global information induced redundancy and thus resulting in an invariant response, that is, the local color nature of the true image assumed by the Gaussian scale mixture model. In summary, the reason why DO cells or other cells in V1 encode the illuminant rather than discount the illuminant is that V1 neurons encoding the illuminant provide useful statistical information of the surroundings that can be exploited by further computations (e.g., DN) for inferring the real invariant characteristics of the image or object (e.g., the intrinsic color) by efficient coding.

\section{Methods}\label{sec2}

\subsection*{Related V1 neural models}

Among others, Burg et al. \cite{burg2021learning} developed several approximate models of V1 neurons by testing various network structures. Their modeling work clearly showed that the classical linear subunit model of V1 equipped with nonlinear divisive normalization could effectively improve the predication accuracy of the model output to the V1 neuron responses to natural images. Our modeling framework is roughly based on Burg et al. \cite{burg2021learning} but with the following key differences and innovations.

%1. The difference from the DN model exploited in Burg et al \cite{burg2021learning} is that 

%Furthermore, the size of input color image is resized as $120\times120\times3$ for computation, which we assume that the resized image approximates the image-computable processing of visual front-end information such as retina and LGN. Hence, the output of revised model is the estimation of illuminant represented by an image-computable quantity (a triplet color value).   

1. In Burg's DN model, the decoding stage uses a spike nonlinear production model based on the Poisson distribution \cite{cadena2019deep}. In contrast, we directly perform a simple convolution on the responses of the DN and then directly use two fully connected layers to decode the color of the light source as the model's estimate. The model's theoretical advantage lies in its utilization of a straightforward linear weighted summation method to combine the output of the RF layer and DN layer. By avoiding the inclusion of additional non-linear operations, the model achieves simplicity and ease of analysis for both the RF and DN layers.

2. The input to our revised model is a color image with a color-cast captured in the real world. The output of the revised model is the estimation of the illuminant represented by an image-computable quantity (a triplet color value). In contrast, a customized gray image is adopted as the visual stimulus to activate the spike responses of Burg's model. The advantages of this approach are that the modified model can effectively utilize both color and luminance information for inference.

3. Our model uses the angle error that measures the cosine similarity between the model's light source color estimate and the real light source color recorded by the camera as the objective function to constrain the training of the V1 neural model. In contrast, log likelihood is used by the DN model as the objective function to fit the neural spike responses to the natural image. By doing so, the model gains the advantage of being able to make predictions based on specific visual task requirements, instead of simply responding indiscriminately to natural image stimuli.

4. We did not perform special processing on the input and output, such as the data normalization adopted by \cite{burg2021learning}. We let the model naturally predict the color of the light source from the input image to see whether the model can use the principle of the triple cross-training strategy to accurately estimate the light source color.

%In summary, the biggest difference is that the original model \cite{burg2021learning} is used to estimate neuron spike patterns directly from the customized gray image without any low-level or high-level tasks, but our model estimates the physical variable from the recorded color image under a constraint of light source color estimate.
\subsection*{The primary structure of the proposed V1 neural models and loss functions for IP}

% Place figure captions after the first paragraph in which they are cited.
\begin{figure*}[!h]
	\includegraphics[angle=0,width=1\textwidth]{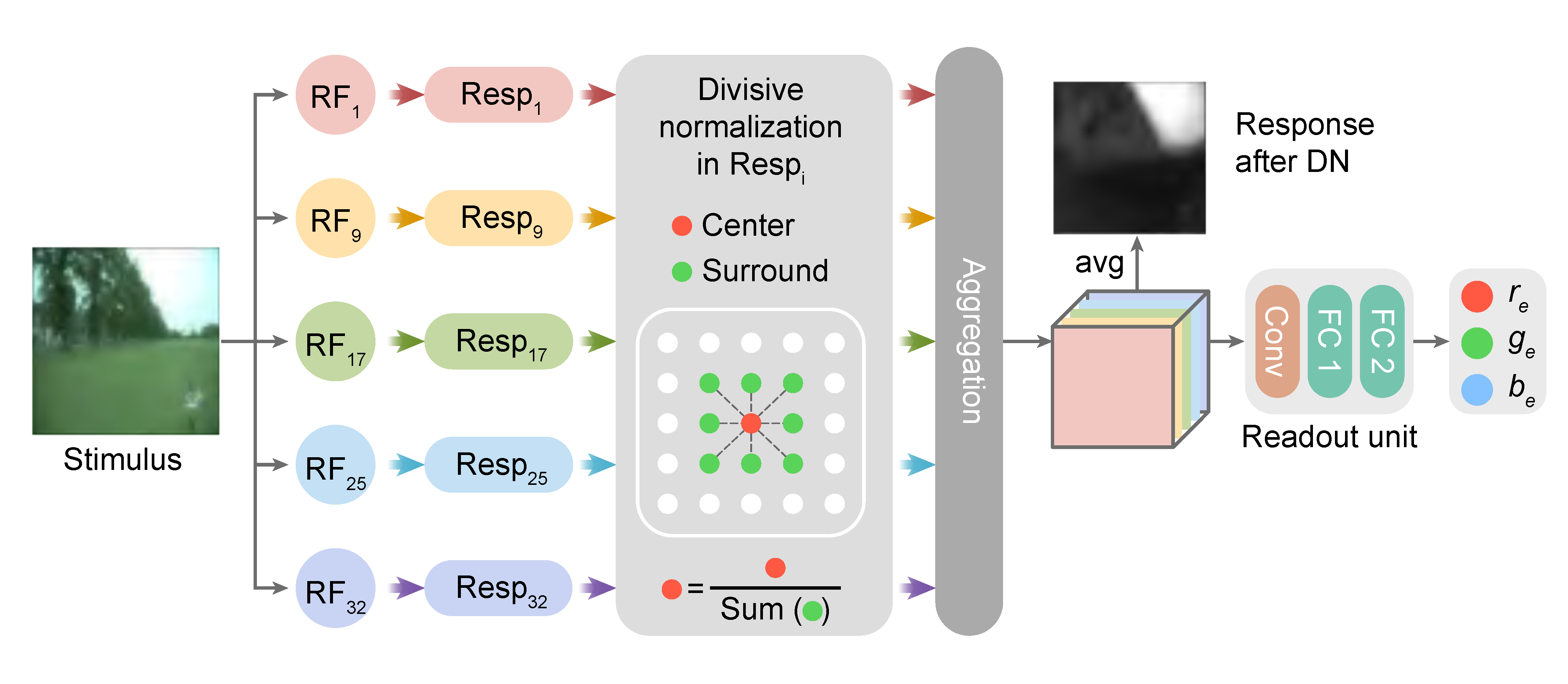}
	\caption{%{\bf Bold the figure title.}
		Overview of the proposed model for IP. This model takes a color image as input and predicts the illuminant color for this scene. Responses after the divisive normalization (DN) are obtained by averaging the response maps of 32 activations of the DN subunits (e.g., channels). %Both images are Gamma corrected to increase the luminance and contrast for better visualization.
	}
	\label{Figure1}
\end{figure*}

%In the following, we will deliberate the details of the proposed V1 model 
As shown in Figure. \ref{Figure1}, our V1 neuron model (i.e., DNM) consists of three parts: the subunit component functioning to linearly filter the input color image, the divisive normalization (DN) component functioning to modulate the signals from the surround RF to the signals of the center through the gain control mechanism, and the readout component functioning to estimate the illuminant by a convolution layer with the size of a $3\!\times\!3$ kernel, followed by two fully connected layers (FC1 and FC2). The fully connected layers model the light source estimate by transforming and weighting the sum of the firing rates representing specific information about the local image patches of different DN subunits \cite{coen2015flexible}. Two fully connected layers fit the activations of the DN subunits to the corresponding illuminant measurement and roughly satisfy biological plausibility \cite{schrimpf2021neural,pogodin2021towards}.

To evaluate the role of each component of the V1 model in predicting illuminant, we build three additional variants of the V1 model. The first one (i.e., Subunit model) is a subunit V1 model consisting of only the linear convolution part shown in Figure. \ref{Figure1}. By comparing these two V1 models, we can determine whether the standard DN contributes to IP or not. The second variant (i.e., DNM with surround) of the V1 model for comparison is a divisive model incorporating a surround unit with the size of $7\!\times\!7$. The difference between the standard DN model and this DN model with surround lies in whether or not the surround signals are included in the divisive normalization. The last variant (i.e., CNN) for comparison is a deep learning model with four linear convolution layers \cite{cadena2019deep,burg2021learning}. 

In other words, we only modify the linear and nonlinear parts of the four V1 models to see whether the different linear and nonlinear parts can confidently learn the meaningful RFs and regions under the only constraint of IP. Finally, we need to explain that although the V1 model is mainly aimed at simulating simple cells in V1, there are new experiments supporting that DO cells also have a similar spatial RF structure and distribution as simple cells \cite{johnson2001spatial,de2021spatial}. Therefore, it is reasonable to use DN in our V1 model.

The primary loss for training all four V1 models is the angular error defined loss written as
\begin{eqnarray}
\label{eq:schemeP1}
Angular \_ loss =cos^{-1}\Big(\big(\vec E_{e}\!\cdot\!\vec E_{t}\big)/\big(\big\| \vec E_{e}\big\|\!\cdot\!\big\| \vec E_{t}\big\|\big)\Big),
\end{eqnarray}
Where $\vec E_{e}\!=\!(r_{e},g_{e},b_{e})$ and $\vec E_{t}\!=\!(r_{t},g_{t},b_{t})$ are respectively the predicted illuminant by the V1 model from the input color image and the true light source color, and $\big(\vec E_{e}\!\cdot\!\vec E_{t}\big)$ is their dot product and $\big\|\!\cdot\!\big\|$ represents the Euclidean norm \cite{gijsenij2011computational}. %The angular error is usually used in computational illuminant estimation to evaluate the performance of various CC models \cite{gijsenij2011computational}. 
Besides the angular loss as the constraint, two regularization terms are also added into the optimization process. The smoothness and sparse prior constraints on the predicted RFs are defined as follows:
\begin{eqnarray}
\label{eq:schemeP2}
Regular \_ loss =\sqrt{\sum{(R\!F\otimes f)}}+\sqrt{\sum{R\!F^{2}}},
\end{eqnarray}
where the first term is the smooth loss and the second term is the sparsity loss on the learned filter kernels $R\!F$, $f$ indicates a Laplace filter with the size of $3\!\times\!3$ \cite{burg2021learning}. Therefore, the final total loss is written as:
\begin{eqnarray}
\label{eq:schemeP3}
Total \_ loss =Angular \_ loss +Regular \_ loss,
\end{eqnarray}

For comparison, we also design a loss that guides the V1 model to learn to discount the illuminant. However, what responses of the V1 neuron should be encoded to discount the illuminant are not specifically deliberated in previous theories \cite{conway2010advances,gegenfurtner2003cortical}. Therefore, we propose the average edge responses of a color-corrected image as the objective to constrain the V1 model to learn to discount the illuminant or perform illuminant-independent responses. The rationale behind this concept is to enable the V1 model to learn predicting image responses without color-bias, thereby partially realizing the previous hypothesis that V1 neurons discount the illuminant. Furthermore, numerous studies have indicated that V1 neurons exhibit robust responses to image edges, thus the V1 model is also trained to predict image edges.
\begin{eqnarray}
	\label{eq:schemeP1}
	Discount \_ loss =cos^{-1}\Big(\big(\vec A_{e}\!\cdot\!\vec A_{t}\big)/\big(\big\| \vec A_{e}\big\|\!\cdot\!\big\| \vec A_{t}\big\|\big)\Big),
\end{eqnarray}
In this task of discounting the illuminant, $\vec A_{e}$ in Equation (4) is the predicted average edge response by the V1 model from the input color image. $\vec A_{t}$ is the average edge response of a color-corrected image rendered under a white light source. 
%\subsection*{Benchmark datasets and experimental settings}

Based on threefold cross-validation, the four V1 neural model-based predictors were evaluated on two typical datasets with various indoor and outdoor scenes \cite{banic2017unsupervised,hemrit2018rehabilitating} for IP. To generally evaluate the accuracy and robustness of IP for the V1 models, we adopted the statistics of the median and mean calculated on the results of each model. The pairwise Wilcoxon signed rank test was used on the results of the validation accuracy to statistically measure the performance difference between two models.

\subsection*{ACKNOWLEDGMENTS}
We would like to express our gratitude
to Zongyi Zhan for his assistance with code debugging, and to
LetPub for their help with polishing the figures. This work was initially completed in 2022 and subsequently presented as a poster and results at the 2023 Systems Vision Science Summer School in T\"ubingen, Germany, and the 
2024 Chinese Vision Science Conference, Guangzhou, China.

\newpage
%\backmatter

{\centering\section*{Supplementary Information}}

\subsection*{More information for the proposed V1 model }
The basic model of a V1 neuron consists of a linear unit, a nonlinear unit, and a readout unit with nonlinear output  \cite{burg2021learning,carandini2005we,heeger1992normalization,solomon2014moving,gao2021explaining,snow2016specificity,coen2015flexible}. For the linear unit, the input image is convolved with the linear kernels or RFs of the neurons to obtain the responses of the linear unit. Then, the responses of each neuron in the linear unit are subjected to a non-linear processing, which primarily implements the divisive normalization (DN) operation as experimentally observed in V1 \cite{heeger1992normalization,solomon2014moving,gao2021explaining,snow2016specificity,coen2015flexible}.

Burg et al. \cite{burg2021learning} found that a three-layer CNN  model can better fit the V1 neuron responses to natural images than the standard V1 model with DN. However, the key difference is that the subunit model of V1 equipped with nonlinear DN can not only obtain good prediction performance to V1's neuron spikes but also provide understandable explanations to some typical V1 phenomena, such as cross-orientation inhibition. DN indicates a specific neural computing strategy, which integrates the signals of the central RF and its surround RF and realizes signal modulation of the surround to the center through the gain control mechanism \cite{heeger1992normalization,solomon2014moving,gao2021explaining,snow2016specificity,coen2015flexible}. DN can explain many nonlinearity phenomena observed in V1 neuron responses, such as the spatial nonspecific suppression and temporal adaptation phenomena observed in many experiments \cite{gao2021explaining,snow2016specificity}. Recent work has shown that DN is a canonical computation in the sensory system that can be utilized in deep neural networks to improve the performance of image classification and reduce the redundancy of outputs \cite{miller2021divisive,pan2021brain,veerabadran2021bio}.

Burg et al. also built three variants of the V1 model for comparison. The only difference between the V1 model with DN and the subunit V1 model is that there is no DN operation in the subunit V1 model. The second V1 model for comparison is a divisive model incorporating a surround unit, which integrates both the spatial signals from the surround neurons and the signals from different channels to normalize the output of the linear unit. The last V1 model for comparison is a deep learning model with three linear convolution layers because this simple convolution network performs best during spike prediction compared with the deeper convolution neural networks \cite{burg2021learning,cadena2019deep}.
%Burg et al \cite{burg2021learning} also show that a three-layer CNN model can do the better fit to V1 neuron responses to natural images than the standard V1 model with divisive normalization. However, the key difference is that the subunit model of V1 equipped with the nonlinear divisive normalization not only can obtain the good prediction performance to V1's neuron spikes, but also show the understandable explanations to some typical V1 phenomena such as cross-orientation inhibition in a silicon experiment. 
%The divisive normalization can explain many nonlinearity phenomena observed in V1 neuron responses such as the spatial nonspecific suppression and temporal adaptation phenomena observed in many experiments \cite{solomon2014moving,gao2021explaining,snow2016specificity,coen2015flexible}. The key structure difference between our V1 model for IP from natural color image and the V1 model \cite{burg2021learning} for fitting spikes of V1 neurons responding to customized gray image is that we replace the original readout part consisting of output nonlinearity based spike production with a convolution layer plus two fully connected layers (e.g., FC1 and FC2) as shown in Figure. \ref{Figure1}. 
The key structural difference between our V1 model for IP from natural color images and the V1 model \cite{burg2021learning} for fitting spikes of V1 neurons responding to customized gray images is that we replace the original readout part consisting of output nonlinearity-based spike production with a convolution layer plus two fully connected layers (FC1 and FC2), as shown in Figure. \ref{Figure1}. The key point here is that we fix the readout part of the four V1 models to facilitate fair comparison. One of the novelties of our paper is that previous work on RF analysis is based on high-level constraints such as object classification and recognition, which seems to be far away from CC \cite{yamins2016using}, in contrast our approach represents the first work based on the constraint of light source color estimation.

In summary, the biggest difference is that the original model estimates the neural spike patterns directly from the customized gray image without any low-level or high-level tasks, but our model estimates the physical variable from the recorded color image under the constraint of the light source color estimate. 
\subsection*{Benchmark datasets and experimental settings}
We adopted two benchmark datasets containing various indoor and outdoor scenes \cite{banic2017unsupervised,hemrit2018rehabilitating} to train the proposed V1 model for the IP task. Both the datasets have been commonly used in computational CC evaluation \cite{gijsenij2011computational}. In specific, the REC dataset contains 568 highly dynamic and high-quality images captured by two cameras \cite{hemrit2018rehabilitating}. The Cube+ dataset contains 1708 high-quality images captured with one camera, and the provided ground truth illuminant for each image is computed from a gray triangle attached to the camera \cite{banic2017unsupervised}.

To test our four V1 neural models on the Cube+ dataset, we adopted threefold cross-validation for evaluation. Specifically, we randomly divided the dataset into three folds. Next, the four V1 models were trained on each two-folds to obtain the learned parameters of the V1 models. Then, the trained V1 models were evaluated on the third fold of the divided dataset for IP performance validation. We repeated 20 times for each fold for the test and let the V1 models be trained from scratch for each fold. By this, we obtained 20 results of the median and mean on each test dataset for every run of the experiment independently. Then, there were three test folds according to the form of the threefold cross-validation, and in total, we obtained 60 results of the median and mean for each V1 model.

To evaluate the accuracy and robustness of the V1 models, we adopted the statistics of the median and mean calculated on the 60 results of each model. The median statistic is a good indicator to measure the accuracy of a model, because it reduces the influence of outliers. In contrast, the mean statistic can better measure the robustness of a model's prediction accuracy \cite{gijsenij2011computational}.

The strategy for testing the four V1 models on the REC dataset is similar to that for testing the four V1 models on the Cube+ dataset, except that we repeated 40 times on each fold for the test because the number of images in the REC dataset is roughly one-third of that in the Cube+ dataset. Thus, on the REC dataset, we obtained 120 results of the median and mean for each V1 model. We adopted fixed threefold cross-validation for evaluation using the divided strategy provided by the REC dataset.

The training details for a V1 model are as follows. A color image as the stimulus is fed to the model, which is processed in turn by the subunit unit, the DN unit, and the readout unit to estimate an illuminant vector with the size of $1\!\times \!3$. Then, an angular error loss is computed and used to automatically calculate the gradients for updating the parameters of the model. The hardware platform is an NVIDIA GeForce 1080 GPU and Intel I7-6700K CPU with Cuda 10.0.130 and Cudnn 7.6.5. We adopted Adam \cite{kingma2014adam} with the learning rate as 0.001 and mini-batch size as 1 for training. We used Tensorflow 1.15 \cite{abadi2016tensorflow,burg2021learning} combined with Python 3.7 to implement the model training and testing. The training and testing settings for the two benchmark datasets are the same as for the four V1 models.

After training on each benchmark dataset, we directly visualize the trained convolution kernels, as shown in Figure. \ref{Figure3} in the main text of each model because the convolution kernels are the only linear filtering operation (e.g., the RF) before the nonlinear operation such as DN, which is similar to the initial signal processing of the V1 neuron. The information is first linearly filtered and processed by the RF of the V1 neurons and then goes to the next step of nonlinear processing. The weights of the linear RF directly reflect the basic information processing property of the V1 neuron. We also visualize the nonlinear part of the V1 neuron, that is, the trained responses of the DN to an input color image. Responses after DN were obtained by averaging the response maps of 32 activations of the DN subunits, as shown in Figure. \ref{Figure11} in the main text. The responses of the nonlinear DN may reflect how subsequent nonlinear operations of a V1 neuron use the information extracted by the previous linear RF processing.
\subsection*{Singular value decomposition}
We adopted the singular value decomposition technique (SVD) as used by De and Horwitz \cite{de2021spatial,horwitz2005paucity} to decompose the RFs into color weights and spatial weights. Based on SVD decomposition, we classified the learned RFs into three types: simple cells for luminance response and red-green and blue-yellow DO cells for color responses according to the taxonomy in the neuroscience literature for comparison. 

Since all of the four V1 models obtained good IP performance and learned the meaningful RFs, we pooled all trained RFs of each V1 model for quantitative analysis without any pre-screening. We want to see whether the models can learn similar RFs in V1 only under the task of IP. We first used SVD to decompose each color RF, as shown in Figure. \ref{Figure3} and Figure. \ref{Figure13} in the main text, into the eigenvalues and eigenvectors. Then, the first row and column eigenvectors were defined respectively as the decomposed color weights and the spatial weights. The color weights represent the weight of each color channel (R, G, B), and its sign and magnitude indicate the contributions from each color channel. Thus, we classify the model neurons into the color-defined type and the luminance-defined type. The decomposed spatial weighting function can represent the spatial RF structure of a processed model neuron. The only difference between our SVD method and the SVD method adopted by De and Horwitz \cite{de2021spatial} is that we did not transform the color weights into the LMS space because we did not know the three phosphor emission spectra of the monitor.

\begin{figure*}[!h]
	\includegraphics[angle=0,width=0.5\textwidth]{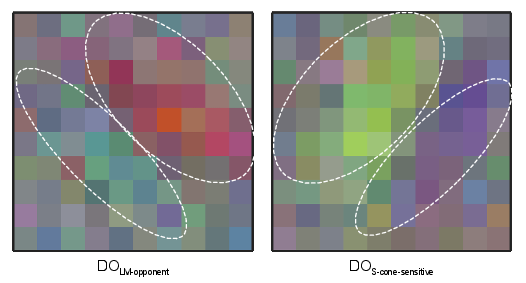}
	\caption{%{\bf Bold the figure title.}
		RFs  of two example V1 cells that were recovered using the spike-triggered average. The white dotted lines overlapping on the figure roughly indicate the subunits of the RF. Adapted from Figure.3C and Figure.3E of De and Horwitz \cite{de2021spatial}.}
	\label{Figure14}
\end{figure*}
\begin{figure*}[!h]
	\includegraphics[angle=0,width=1\textwidth]{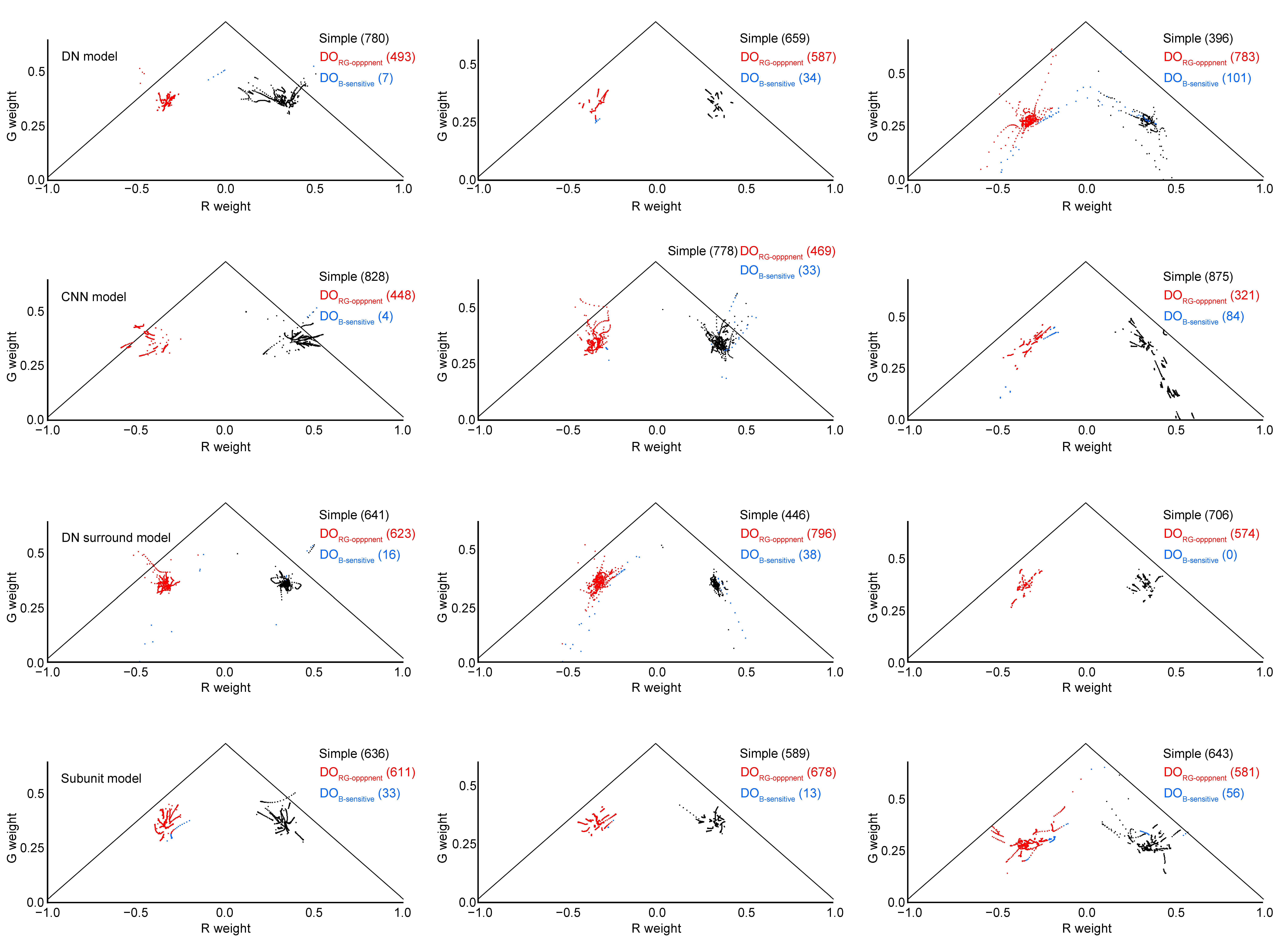}
	\caption{%{\bf Bold the figure title.}
		Cell category distributions of the trained RFs based on their color weight function calculated on the REC dataset \cite{hemrit2018rehabilitating}. From top to bottom: the DN model, the CNN model, the DN model with surround, and the subunit model. From left to right: fold 1, fold 2, and fold 3, where each fold contains 40 results and each epoch has 32 trained kernels (i.e., the RF of the model neurons). Therefore, each subfigure consists of 1280 model neurons. The RFs are classified as simple cells (labeled with black points) if their color weights of the R and G channels are both positive and account for a large number of the weights. The RFs are classified as DO cells with red-green color opponency (labeled with red points) if their color weights of the R and G channels are opposite and account for a large number of weights. The RFs are classified as the DO cells with blue-sensitive color opponency (labeled with blue points) if their colors weights of the R and G channels are opposite and the corresponding B weight accounts for a large number of weights or their color weights of the R and G channels are both positive but their corresponding B weight is negative and accounts for a large number of weights. The criteria of the normalized weight classification are roughly according to De and Horwitz \cite{de2021spatial}.}
	\label{Figure5_RCC}
\end{figure*}

\subsection*{Weight randomization}

We conducted an ablation experiment to determine which kinds of V1 neurons perform better in IP. Based on the neuron classification results, we randomized the learned weights of the simple cells and kept the learned weights of the DO cells unchanged. Then, we used this model only with the DO cells to predict the performance of the light source color estimation on the test dataset. Similarly, we also randomized the learned weights of the DO cells and kept the learned weights of the simple cells unchanged. Then, we used this model only with the simple cells to predict the performance of the light source color estimation on the same test dataset. The full model with both the DO cells and simple cells was adopted as the baseline for comparison.

\subsection*{Mathematical basis of V1 DO cells encoding the illuminant for reducing redundancy} 
A Gaussian scale mixture model (GSM) can be used to describe the nonlinear responses of V1 neurons \cite{coen2015flexible} to a natural image. In GSM, the responses of a V1 neuron are determined by Bayesian inference processing. GSM assumes that the V1 neurons calculate an optimal estimate of the local image property within the RF by eliminating the redundancy between the center neuron's representation and the surround neuron's representation. Basically, GSM defines the center neuron's responses $C$ and surround neuron's responses $N$ after a bank of V1-like RFs as follows. Without loss of generality, here we assume that the neuron responses of $C$ and $N$ are from DO neurons in V1.     
    
\begin{align}\left\{\begin{aligned}
		C&=\nu c\\
		N&=\nu n
	\end{aligned}\right.\end{align}
where $\nu$ indicates a random positive variable named mixer, $c$ and $n$ are two local random variables that meet the Gaussian distribution. Essentially, GSM treats the responses of V1 neurons to a color image as a generative model process, where $C$
and $N$ are generated through multiplying a Gaussian variable $c$ or $n$ with a random variable $\nu$. Intuitively, the variable $\nu$ usually represents the global property of an image patch that causes the non-linear dependency between the responses of the center neuron and the responses of the surround neuron. The Gaussian variables $c$ and $n$ represent the strengths of the local image features in the RF of the center neuron and the RF of the surround neuron that the V1 neurons should finally encode. A hallmark feature when applying V1-like filters to a natural image is that there is typical bowtie dependence as captured by Equation (5), which exists in many natural signals, such as sound, NMR signal, and natural image \cite{schwartz2001natural}.

However, the global property of an image patch is not clearly known, and it may be the global contrast or other global information across the image. For a color image, the global information of an image patch should contain the global illuminant since the illuminant is usually evenly distributed for each raw pixel or the response of a pixel after a linear filtering such as the DO cell. Hence, we can treat the variable $\nu$ as the global color information representing the illuminant and the Gaussian variables $c$ and $n$ as the local image features such as the intrinsic local color property of an image patch. To reduce the redundancy caused by the global property and achieve efficient coding according to the efficient coding theory \cite{dayan2001theoretical}, GSM mathematically estimates the neuron's average responses, equivalent to using divisive normalization \cite{coen2012cortical}.

\begin{equation}
\begin{aligned}
\overline{c}\approx\frac{C}{\sqrt{(C,N)^{T}(\Sigma^{}_{cn})^{-1}(C,N)}}, 
\end{aligned}
\vspace{-0.3em}
\end{equation}  

where $\overline{c}$ is the estimated expectation of the local Gaussian variable $c$, which essentially represents the features that the center neuron should encode. $N$  represents the surround neuron's responses to a color image, where we simply abstract the surround as consisting of some V1 neurons. $(\Sigma^{}_{cn})^{-1}$ indicates the covariance matrix that fully describes the dependency between the local Gaussian variables $c$ and $n$.
Equation (6) can be further simplified as 

\begin{equation}
\begin{aligned}
\overline{c}\approx\frac{C}{\sqrt{(\Sigma^{}_{cn})^{-1}C^{2}+2(\Sigma^{}_{cn})^{-1}CN+(\Sigma^{}_{cn})^{-1}N^{2}}}, 
\end{aligned}
\vspace{-0.3em}
\end{equation}  

By substituting Equation (5) into Equation (7), we obtain

\begin{equation}
\begin{aligned}
\overline{c}\approx\frac{\nu c}{\sqrt{(\Sigma^{}_{cn})^{-1}(\nu c)^{2}+2(\Sigma^{}_{cn})^{-1}(\nu c)(\nu n)+(\Sigma^{}_{cn})^{-1}(\nu n)^{2}}}, 
\end{aligned}
\vspace{-0.3em}
\end{equation}  

where $\nu$  represents the global color information representing the illuminant.  
We can see that divisive normalization cancels out the shared variable $\nu$ in the numerator and denominator of Equation (8) and thus obtains an independent estimation of the local Gaussian variable $c$. The responses of Equation (8) are independent of the light source color, which are the assumed responses of V1 to the local intrinsic properties of an image patch. It should be noted that this inference is based on the prerequisite that the neural responses of $C$ and $N$  from the DO neurons in V1 can be described by Equation (5). In other words, the neural responses of $C$ and $N$  from the DO neurons in V1 coherently encode the illuminant through a generative model. 

However, if the responses of $C$ and $N$ from the DO neurons in V1 instead discount the illuminant such as the responses of $C$ in Equation (5) does not rely on the global variable $\nu$ or the responses of $N$  in equation (5) does not rely on the global variable $\nu$, the result of divisive normalization cannot completely eliminate $\nu$ in the numerator and denominator, and it still carries $\nu$. Therefore, under the single-illuminant condition, the responses of the V1 neurons and specifically those of the DO cells should encode the illuminant either after the filtering stage or after the DN stage. In other words, the responses of the V1 neurons and specifically those of the DO cells encode the illuminant, which is a global property representing redundancy that should be satisfied if the cortex is optimized to obtain the efficient representation by reducing the redundancy deeply rooted in a rich literature \cite{dayan2001theoretical}.

%Hence, under a single-illuminant condition the responses of V1 neurons specifically the double-opponent cells should encode the illuminant no matter after the filter's stage or after the DN stage. In other words, the responses of V1 neurons specifically the double-opponent cells encode the illuminant that is a global property representing redundancy should be satisfied if the cortex is optimized to obtain the efficient representation by reducing the redundancy deeply rooted in a rich literature \cite{dayan2001theoretical} is established.  

%However, why the DO cells with special RF of both spatial-opponency and color-opponency are good at encoding the illuminant? 

\subsection*{Extended discussion and conclusion}\label{5}
Our objective is to investigate whether the trained V1 models consisting of linear and nonlinear components can learn to predict acceptable illuminant effects from a dataset that is not used for training. Furthermore, we examine whether the electrophysiologically based V1 neuron model can learn meaningful RF structures as shown in recent electrophysiological experiments only under a low-level constraint. The findings show that under the only constraint of light source color prediction, the function of the learned DO V1 neurons is indeed to estimate or encode the color of the light source, regardless of the model implementation (i.e., subunit model, divisive model, or CNN model). Our work further clarifies the visual perception and image processing functions of DO cells in V1 and directly confirms that some neurons in V1 show illuminant-dependent responses. %DO cells or other cells in V1 encoding the illuminant can help achieve the efficient coding that the DN uses with the surrounding statistical information encoded by V1 neural processing to infer the real invariant characteristics of the image or object (e.g., the intrinsic color).
DO cells, as well as other V1 cells responsible for encoding illuminant, are instrumental in achieving efficient coding. By processing the surrounding statistical information encoded by V1 neurons, the DN can then infer the invariant characteristics of an image or object, such as its intrinsic color.

%\nolinenumbers 
%\bmhead{Acknowledgments}

%Acknowledgments are not compulsory. Where included they should be brief. Grant or contribution numbers may be acknowledged.

%Please refer to Journal-level guidance for any specific requirements.


\bibliography{doublePMAI}% common bib file


\end{document}